\documentclass[twocolumn,pra,superscriptaddress,showpacs,preprintnumbers,amsmath,amssymb]{revtex4}
\usepackage[ansinew,latin1]{inputenc}
\usepackage{array}
\usepackage{amssymb}
\usepackage{amsmath}
\usepackage{graphicx}
\usepackage{xcolor}
\usepackage[english]{babel}

\begin{document}

\title{Progress towards Bell-type polarization experiment with thermal neutrons}

\author{M. Iannuzzi}
\affiliation{Dipartimento di Fisica, Universit\`a di Roma ``Tor Vergata", I-00133 Roma, Italy}
\author{R. Messi}
\affiliation{Dipartimento di Fisica, Universit\`a di Roma ``Tor Vergata", I-00133 Roma, Italy}
\affiliation{INFN Sezione ``Roma Tor Vergata", I-00133 Roma, Italy}
\author{D. Moricciani}
\email{dario.moricciani@roma2.infn.it}
\thanks{Corresponding author}
\affiliation{INFN Sezione ``Roma Tor Vergata", I-00133 Roma, Italy}
\author{A. Orecchini}
\affiliation{Dipartimento di Fisica e Geologia, Universit\`{a} di Perugia, I-06123 Perugia, Italy} 
\affiliation{Institut Laue-Langevin, F-38042 Grenoble, France}
\author{F. Sacchetti}
\affiliation{Dipartimento di Fisica e Geologia, Universit\`{a} di Perugia, I-06123 Perugia, Italy} 

\date{\today}

\begin{abstract}
Experimental tests of Bell-type  %\cite{bell} 
inequalities  distinguishing between quantum 
mechanics and local realistic theories remain of considerable interest if performed on massive 
particles, for which no conclusive result has yet been obtained. Only two-particle 
experiments may specifically test the concept of spatial nonlocality in quantum theory,
whereas single-particle experiments may generally test the concept of quantum noncontextuality. 
Here we have performed the first Bell-type experiment with a beam of thermal-neutron pairs in 
the singlet state of spin, as originally suggested by J.S. Bell. These measurements confirm  
the quantum-theoretical predictions, in agreement with the results of the well-known 
polarization experiments carried out on optical photons years ago.
\end{abstract}


\maketitle

Since the starting point of Bell's theorem on non-locality of the quantum theory in 1965 \cite{bell}, 
theoretical speculations and search for experimental verifications of various inequalities derived 
from the theorem have been pursued over the years. Most of the performed experiments   
were based on correlations between the polarization of pairs of photons, in particular high-energy 
photons produced by positronium annihilation \cite{bruno}, or optical photons emitted by atomic 
cascade \cite{freeman,aspect} or by parametric down-conversion \cite{weihs}. One test was 
also done on massive particles  to measure the spin correlation in proton pairs prepared in 
low energy proton-proton scattering \cite{lamehi}. These early experiments on positronium 
annihilation and proton-proton scattering did not give conclusive results, mainly because of 
the lack of efficient linear polarization filters both for protons and high-energy photons, and 
because of other difficulties related to the apparatus.
One more recent and complex test performed on massive particles, pairs of entangled $^9$Be$^+$
ions, appears as conceptually questionable because the manipulations of the two particles 
of a pair were not done individually and independently of each other, as would be required to 
verify spatial non locality \cite{rowe}.
The extensive tests working with atomic cascades producing pairs of optical photons have been 
the most successful (together with those completed afterward by using parametric down-conversion), 
mainly because of the availability of very efficient polarization filters and an efficient technique for 
the manipulation of visible light beams. The experimental limitations of such experiments 
are mainly related to the relatively low quantum efficiency of the detectors and the difficulty to select 
and detect both members of a photon pair, because the variable recoil momentum of the emitting 
atom diminishes their direction correlation. However, with a supplementary assumption justifying 
that the detected photon pairs constitute a statistically representative sample of the whole 
ensemble of photons emitted from the source, these experiments confirm the predictions of the 
quantum theory with great accuracy. A detailed and comprehensive analysis of the three classes 
of test experiments performed on Bell's theorem (e.g. atomic cascades, proton-proton scattering, 
positronium decay), is given in \cite{clauser1}.

The above considerations, and particularly the limits of the experiments with massive particles, have 
motivated us to perform a Bell-type experiment working with a collimated beam of thermal neutrons 
prepared in such a way that, within the experimental limitations of the measurement, neutron 
pairs in the singlet state of spin were identified by the detecting system. We recall 
that an experimental test on such a  physical system is particularly interesting since Bell's inequality 
is violated by a large amount in the singlet state. We also recall that beautiful interferometric 
measurements have  been done with single-neutron beams, testing correlations between two degrees 
of freedom of the single particle \cite{hasegawa1,hasegawa2,bartosik}. 
However such experiments are not equivalent to those performed with two-neutron beams, which 
refer to the issue of spatial nonlocality for two separated particles. 
This latter concept is meaningless for a single particle, and the interferometric experiments have 
rather tested the general concept of quantum noncontextuality.

Following the results of the previous measurements on the fermion anti-bunching \cite{iannuzzi1,
iannuzzi2}, the present measurements have been carried out at the Institute Laue Langevin 
(Grenoble, France) by using the primary spectrometer of instrument IN10, which produces a 
monochromatic beam of thermal neutrons from an {\it almost} perfect Si(111) single crystal in 
the almost perfect backscattering configuration. The neutron intensity at the test 
position was $2 \times 10^3$ n/s on a beam size of $3 \times 4$ cm$^2$. 
The energy spread cannot be measured directly but it can be estimated from the monochromator 
geometry to be $\Delta E < 0.02 \, \mu$eV, corresponding to a coherence time $\tau_c = \hbar / 
\Delta E >$ 20 ns at 6.27 \AA.
The distance of the multianode detector from the monochromator was $\sim$ 12 m and $\sim$ 2 m
from the collimator exit; the whole equipment was within the shielding of IN10.

The extraction of a beam of correlated particles in the singlet state can be described as follows.
The nature of the emission of thermal neutrons in the source (the neutron moderator of ILL) 
is Poissonian, so that there is a small but finite probability of having two neutrons within the 
detection time of our apparatus. The beam emitted from the source is monochromatized by 
reflection on a quasi-perfect Si crystal. The beam being spin unpolarized, it is equally likely 
that a neutron pair emitted within the coherence time $\tau_c$ of the monochromator will either occur 
in one of the three triplet states or in the singlet state, i.e. the triplet states will occur  3/4 of the 
time. In a gas of fermions there is the tendency for particles of the same spin to avoid each 
other, a tendency arising from the exchange antisymmetry of the wave function. 
More specifically the two neutrons of a pair emitted in the  triplet state and traveling along 
the long collimators ($\sim$10 m) from the monochromator to the polarizers are in the same 
spatial mode (antisymmetrical spatial wavefunction), so that they avoid each other scattering in 
directions different from the one of the emerging collimated beam, and they will not reach the 
detectors. 
The neutrons of a  pair in the singlet state do not avoid each other traveling along the 
collimators; they will pass through the two separated polarizers, and after them, 
are distinguishable particles that will be detected without exhibiting interference effects. 
Consequently, within a time interval of the order of $\tau_c$, only neutron pairs emitted 
from the source in the singlet state can be detected in our apparatus.
This is the physical effect already measured in \cite{iannuzzi1,iannuzzi2}, and applied 
in the present experiment.
 
\begin{figure}%[hb]
  \centering
  \includegraphics[width=6.2cm,angle=-90]{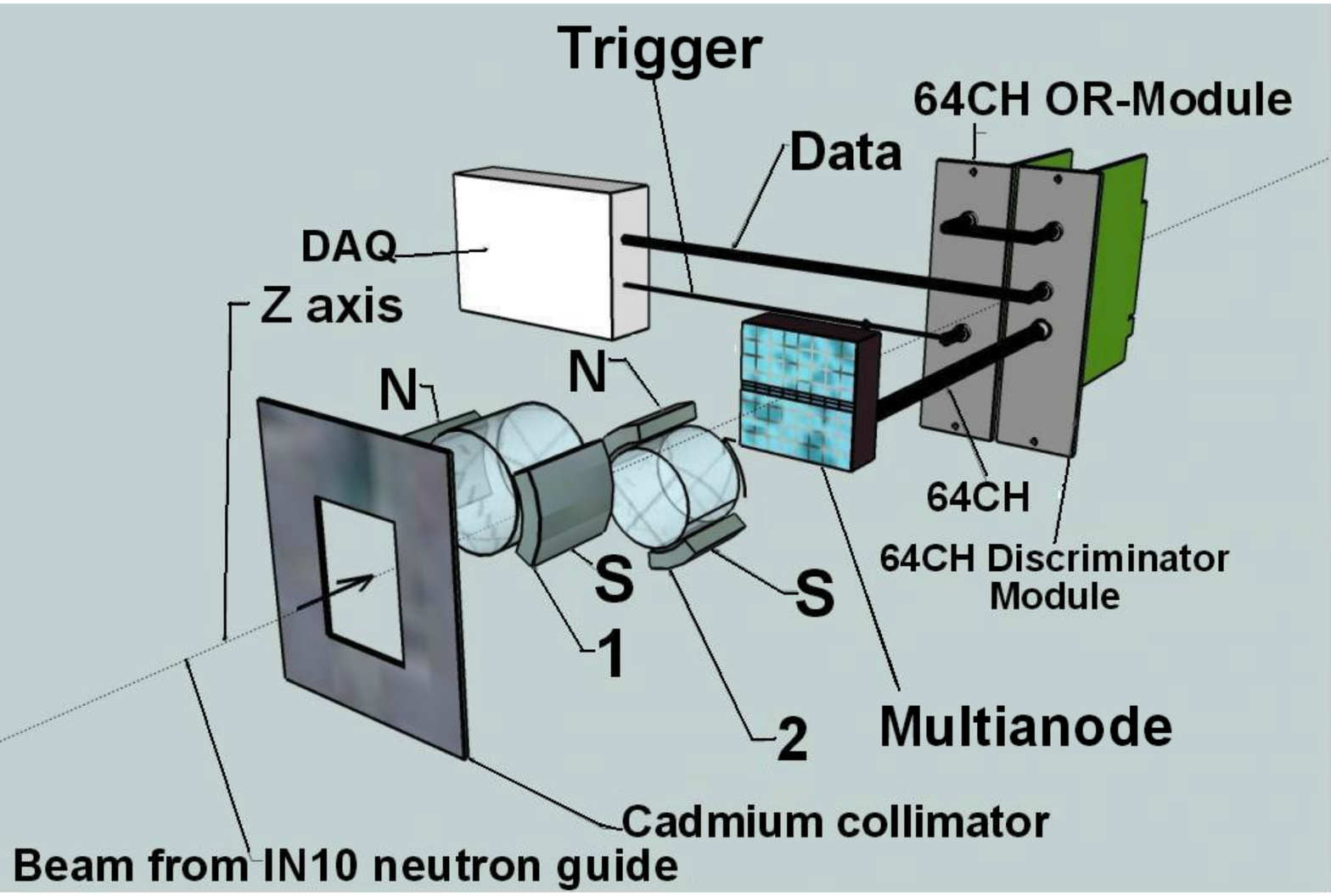}
  \caption{Scheme of the experiment. Distances along the $z$-direction are not on scale. 
  Sizes of the neutron guide and collimating window are respectively 3$\times$3 cm$^2$ and 
  3$\times$4 cm$^2$. 64CH = 64 channels from the multi-anode pixels;
  DAQ = data acquisition system; OR-module is a logical circuit triggering DAQ on 
  receiving a signal from whatever channel of the 64CH discriminator module.
  The polarizers are separated vertically, fronting the two halves of the detectors; erroneous 
  visual impression may arise from the drawing perspective.}
  \label{fig1}
\end{figure}

The conceptual scheme of the experiment was similar to the one described in \cite{iannuzzi2}, with 
the addition of two polarizers (1 and 2 in Fig.\ref{fig1}), each one mounted along two slightly different 
neutron paths from the beam output window.
Both polarizers consisted of a disk, 3.8 cm diameter and 2.6 cm thick, made of Fe$_3$Al 
intermetallic compound and mounted between the poles of a 0.2 T permanent magnet. 
The Fe$_3$Al inter-metallic compound was chosen because the first Bragg edge, 
corresponding to the (111) reflection of its DO$_3$ structure, is at $\lambda$ = 6.67 \AA, 
while the second Bragg edge due to the (200) reflection is at $\lambda$ = 5.78 \AA. So, at 
the working neutron wavelength of 6.67 \AA~the transmission of the disk is due only to the capture 
cross section of Fe and to the (111) Bragg scattering. 
If a polarizer is magnetically saturated, the cross section is large when the $S_z$ 
component of the neutron spin is parallel to the magnetization, and it is small when $S_z$ is 
antiparallel \cite{fe3al}. 
With our 0.2 T magnet the measured transmission coefficients were $\varepsilon^\downarrow$ = 
22.1\% and $\varepsilon^\uparrow$= 8.4\%. In the present apparatus, polarizer 2 
was mounted on a motorized rotation stage and could be rotated 
with respect to polarizer 1 mounted within a fixed horizontal field. 
The relative angle $\vartheta$ of their polarization axes was variable in the range from 
0$^\circ$ to $360^\circ \pm 0.1^\circ$. Since the polarizers can be considered as state 
preparation devices, their position along the neutron path was not important, provided 
that the two neutron paths were identical as it was in our apparatus.

As in \cite{iannuzzi2}, the position-sensitive detector was a Hamamatsu H8500 multi-anode 
photomultiplier having 8$\times$8 anode-pixels with a pixel size of 5.8$\times$5.8 
mm$^2$. The photomultiplier was coupled to a 0.15 mm thick lithium glass scintillator 
($^6$Li 98\% enriched) directly coupled to the anode window with a thin layer of optical 
grease. The scintillator was divided into two different parts with a separation distance of 
$\simeq$ 0.4 mm, so that each part of the scintillator was looking at the neutrons passing 
through either polarizer 1 or polarizer 2, and there was no cross-talk between the two 
detecting areas due to the neutrons impinging on the separation area. 
Specific pixel groupings could be defined during the off-line analysis to 
optimize the view of the upper and lower parts of the neutron beam, so that each of the  
two detectors, $D_1$ and $D_2$, recorded the neutrons passing through one
specific polarizer.  Generally each detector was formed of a row of 4-6 
pixels with total area $\le$ 3.0$\times$0.5 cm$^2$, chosen so as they were illuminated 
only by those neutrons that, filtered respectively by polarizer 1 and polarizer 2, could 
not be intercepted by the Fe magnetic poles of polarizer 2 at any angle $\vartheta$.
The neutron arrival time at each pixel was measured by using an internal 40 MHz clock 
which provided 25 ns time resolution, shorter than both the light decay time of the lithium 
glass scintillator ($\simeq$ 250 ns) and the neutron total travelling time through the 
scintillating sheet ($\simeq$ 240 ns). The data acquisition system monitored events 
during repeated cycles of 10 s each, with an effective duty cycle of about 99.90\%, 
where the dead time due to the VME read out cycle was of the order of 10 ms.

In order to determine the spin correlation of two neutrons, a time stamp was attributed to the 
neutron signals from the two detectors, so that the neutron pair was time correlated 
when two neutrons arrived at the two detectors $D_1$ and $D_2$ with time-stamp 
$t_1 =$start and $t_2 =$stop. In the off-line analysis we could study the 
coincidence rate as function of $\delta=t_2-t_1$, or in other words as a function of 
the virtual spatial separation $v_{\rm ther}\delta$ along the propagation 
direction $z$.
In \cite{iannuzzi1,iannuzzi2} it is also shown that, thanks to the specific characteristics of the 
IN10 primary spectrometers, there is a good neutron correlation when the two detector 
$D_1$ and $D_2$ are separated in the transverse direction up to a couple of centimeters. 
The issue of the existence or amplitude of macroscopic lateral coherence and its possible effects 
on coincidence measurements is controversial. For this reason we prefer to limit ourselves
to report our measured value. We have no conclusive interpretation of this results, and we feel 
that this effect requires further experimental investigation.

According to the above description, the experimental results that we are presenting provide the 
number of  neutrons monitored by $D_2$ at a time $t_2$ delayed 
by $\delta$ with respect to time $t_1$ when the first neutron is detected at $D_1$.  
It is interesting to obtain a calculation of this number from the convolution of the predicted
neutron-correlation function $c(\vartheta,\delta)$, and the response function of the 
detecting system $w(\delta)$. 

The function $w(\delta) = W \exp[-\delta/t_w]$ describes the time broadening of the detecting system 
due to  the scintillator thickness and to the decay time of the emitted light. Actually, from the 
capture probability profile of the neutron along the scintillator thickness, the average capture 
time is about $t_c = 80$ ns with an additional broadening from the light decay curve of the 
order of $t_d = 250$ ns. 
By considering that both effects can be roughly described by an exponential function, 
the total decay time is expected to be $t_w = (1/t_c + 1/t_d)^{-1} =$ 60 ns. 
The choice of the Gaussian form of $c(\vartheta,\delta) = 1 - \alpha(\vartheta) \, \exp[-\delta^2
/(2\tau_c^2)]$ was dictated first by the need of describing reasonably (similarly to what 
was done in \cite{iannuzzi1}) the antibunching effect which produces, for $\delta=0$ and 
for any value of $\vartheta$, a 3/4 depression of the coincidences counts with respect to the 
random coincidences occurring for $\delta\gg\tau_c$; and second by the need 
to add also the simultaneous depression produced by the presence of the polarizers 
on the neutrons pairs in the singlet state (see prediction (\ref{eq7}), recalled later in the text).
The value of the coefficient $\alpha(\vartheta)$ must be $3/4+1/4=1$ for $\vartheta=0^\circ$.
%The number of the coincidence is then given by:
From the convolution $[w*c]c(t')$ of $w$ and $c$, the number of coincidences
is then given by:

\begin{equation} 
C(\vartheta,\delta) = \frac{C(\vartheta,\infty)}{\Delta}\int_\delta^{\delta+\Delta}[w*c](t')dt'
~~~~~~~~\delta>0
\label{eq1}
\end{equation}

\noindent 
where $C(\vartheta,\delta)$ is the number of experimental coincidences as a function of 
$\delta$,  $C(\vartheta,\infty)$ is the number of experimental 
coincidences for  $\delta\gg\tau_c$, $\Delta$ is the coincidence time-window of the order of
$\tau_c$, and $W=1$.  
In Fig. \ref{fig2} the experimental results are compared with the prediction of Eq. ({\ref{eq1}).
Tests taking $\alpha(0)$, $t_w$ and $\tau_c$ as free parameters confirmed our estimates,
and we found $\alpha(0) = 1 \pm 0.1$, $t_w=60\pm 9$ ns and $\tau_c = 78 \pm 10$ ns, in 
agreement with the previous estimates \cite{iannuzzi2}.

\begin{figure}%[hb]
  \centering
  \includegraphics[width=8.6cm]{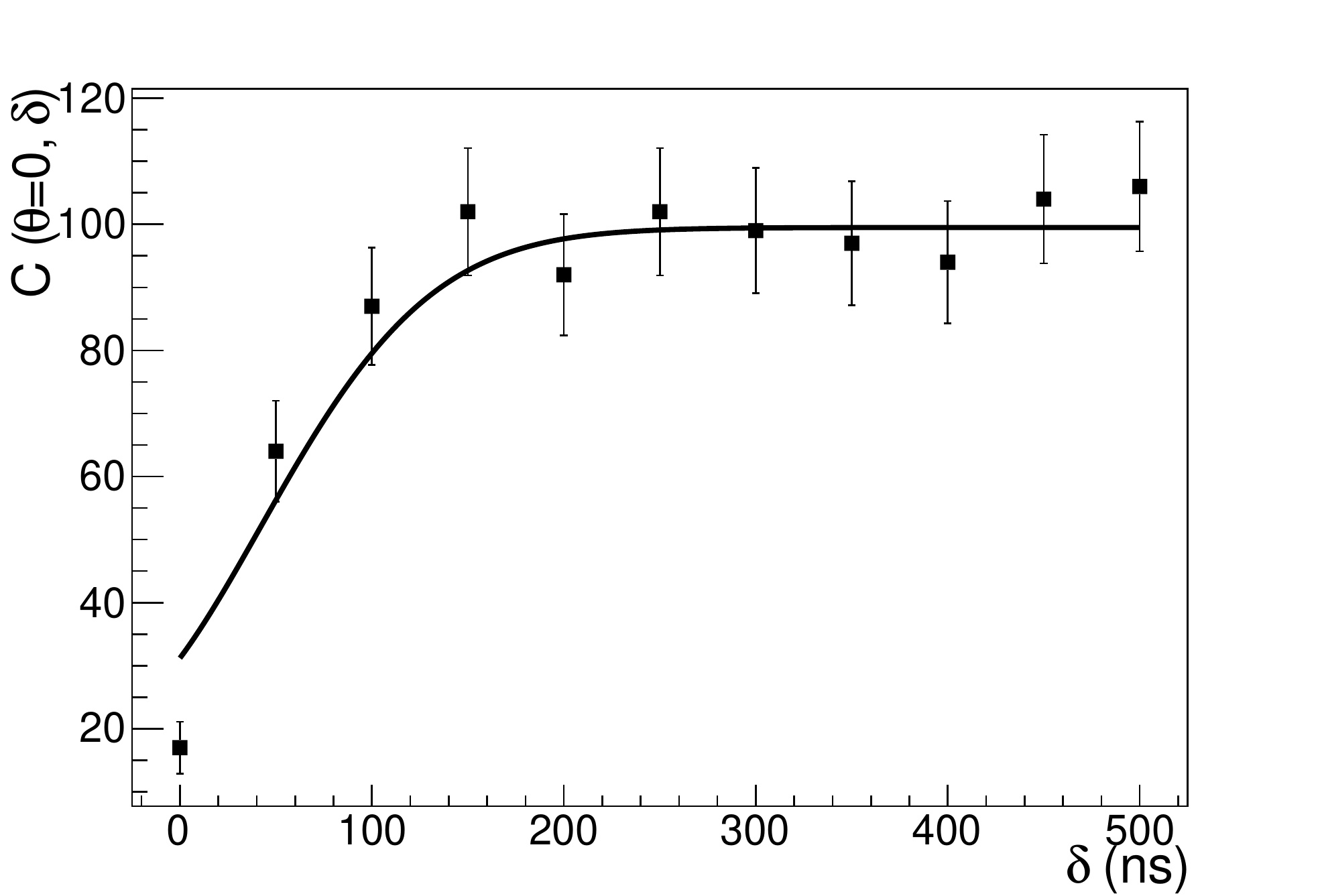}
  \caption{$C(\vartheta=0,\delta)$ is shown as a function of the neutron time-delay $\delta=t_2-t_1$ 
  along the longitudinal $z$-direction; full square = experimental points. Two 4-pixel rows with relative 
  separation 0.4 cm, $\Delta$=150 ns. }
  \label{fig2}
\end{figure}

We can now define the function:

\begin{equation} 
P_{12}(\vartheta) = \frac{C(\vartheta,0)}{C(\vartheta,\infty)}
\label{eq2}
\end{equation}
 
\noindent
which is independent of all possible fluctuations of the beam intensity and other sources
of instrumental uncertainties. It represents the experimental value of the probability of detecting 
the simultaneous occurrence, after transmission through the linear polarization analyzers, 
of the two neutrons of a pair emitted in the singlet state, with respect to all four states of 
the physical system. 
The experimental results obtained for $P_{12}(\vartheta)$ at five different values of $\vartheta$ 
are shown in Fig. \ref{fig3}. The acquisition time of each experimental point was in the range 
18-24 h.

In order to obtain the quantum-theoretical prediction for the outcome of the experiment, 
let us first calculate the probability $P_{12}(\vartheta)$ for perfect analyzers that transmit 
only neutrons with spin antiparallel to the magnetic field. 
During a given time interval, the source emits $N$ neutron pairs. For $N$ large enough, 
we define: $N_i/N = c_i$, $N_i$ being the number of counts at detector-{\it i}; and we also 
recall that $N_i=N/2$ is the number of neutrons transmitted through perfect polarizer {\it i} 
when $\delta = t_2 - t_1 \gg \tau_c$. 
Since the fraction of the neutron pairs emitted in the singlet state and recorded after the 
polarizers  at short time delays is 1/4 of all the pairs, we obtain

\begin{equation}
P_{12} (\vartheta) = \frac{1}{4}\langle \psi_s|Q^\downarrow_1Q^\downarrow_2|\psi_s \rangle = 
\frac{1}{4}\frac{[1-\cos(\vartheta)]}{4} 
\label{eq3}
\end{equation}

\noindent
where $\vert \psi_s \rangle$ is the singlet wave function and $Q^\downarrow_i$ is the operator 
which projects the spin of particle ({\it i}) on the direction of magnetic field %({\it i}) 
with value -$\hbar/2$.
Yet a more realistic analysis, taking account of the limitations of the apparatus, can be adopted 
for comparison with the real experiment. 
We can adapt to result (\ref{eq3}) 
the analysis adopted by Clauser, Horne and Shimony (CHS) \cite{clauser1,clauser2} for pairs of optical 
photons with parallel linear polarization. 
They derive the following prediction: $P_{12}(\vartheta)=c_1c_2[\varepsilon_1^+\varepsilon_2^+ +
\varepsilon_1^-\varepsilon_2^-\cos(2\vartheta)]/4$ and $P_{12}(\infty,\infty) = c_1c_2$ with 
$\varepsilon^+=\varepsilon_i^M+\varepsilon_i^m$, $\varepsilon^-=\varepsilon_i^M-\varepsilon_i^m$, 
where $\varepsilon_i^M$ and $\varepsilon_i^m$ are respectively the maximum and minimum 
transmittance of polarizer $i$. The symbol $\infty$ defines  the absence of a polarizer.

By adapting their approach, particularly the one in \cite{clauser1}, to our system of thermal 
neutrons, we have not performed a direct mathematical derivation of their result, because the 
significative differences to be considered between the two cases have dictated to introduce 
changes in their calculation.
First we recall that a linearly polarized photon propagating along the $z$ axis has polarization 
direction in the $x$-$y$ plane; differently a neutron propagating along the $z$ axis has spin 
components different from 0 only in one $x$-$y$ semiplane. 
We also recall that the optical linear polarizers have the maximum transmittance in direction 
perpendicular to that of minimum transmittance, while thermal neutron polarizers have maximum 
and minimum transmittance in antiparallel directions. 
Second, only coincidences occurring within a time interval $\delta\le\tau_c$ are 
considered in our analysis, i.e coincidences due to pairs of neutrons in the singlet state. 

We must now remark that a neutron polarizer like ours is a linear polarizer with oriented polarization 
axis, so that two of such perfect polarizers ($\varepsilon_1^\downarrow=1$ and $\varepsilon_2^\uparrow=0$)
with parallel orientation of their axes cannot transmit two correlated neutrons in the singlet state (antiparallel 
spins), and coincidences may be observed only in the angular range $90^\circ\le\vartheta\le 180^\circ$.
Consequently, the ideal result (\ref{eq3}) can only be confronted with Eq. (\ref{eq6}), obtained later  
in the text, in this angular range.
However, with real polarizers ($\varepsilon_1^\downarrow=0.221$ and $\varepsilon_2^\uparrow=0.084$ 
in our experiment) with oriented polarization axis, coincidence counts may also occur for $0^\circ\le\vartheta\le 
90^\circ$, but still the probability of their occurrence cannot be confronted with prediction (\ref{eq3}).

As a consequence of such differencies between optical photons and neutrons, in the above CHS equation
$\varepsilon_i^+$ and $\varepsilon_i^-$ have become respectively $\varepsilon_i^\downarrow$ and 
$\varepsilon_i^\uparrow$ in the term proportional to $\cos(\vartheta)$.
In addition, note that the neutron correlation function $\sin^2(\vartheta/2)/2$ [see Eq.(\ref{eq7})] substitutes the 
photon correlation function $\cos^2(\vartheta)/2$; the symbol $\infty$ does not denote the absence 
of a polarizer, it denotes the absence of correlation between to two spins of a pair for $\delta\gg\tau_c$
instead; the beam of neutrons in the singlet state is the fraction 1/4 of the whole ensemble emitted 
from the monochromator.

Following the above considerations, we obtain for $0^\circ \le \vartheta \le 90^\circ$

\begin{equation}
\begin{split} 
& C(\vartheta,0) = \frac{c_1 c_2}{4}\Bigr [\frac{\varepsilon_1^t \varepsilon_2^t - (\varepsilon_1^\downarrow
\varepsilon_2^\uparrow + \varepsilon_1^\uparrow \varepsilon_2^\downarrow) \cos(\vartheta)}{4}\Bigl ]
\\
& C(\vartheta,\infty) = \frac{c_1 c_2\varepsilon_1^t \varepsilon_2^t}{4}
\end{split}
\end{equation}

\noindent
and for $90^\circ \le \vartheta \le 180^\circ$

\begin{equation}
\begin{split}
& C(\vartheta,0) = \frac{c_1 c_2}{4}\Bigr [\frac{\varepsilon_1^t \varepsilon_2^t - (\varepsilon_1^\uparrow 
\varepsilon_2^\uparrow + \varepsilon_1^\downarrow \varepsilon_2^\downarrow) \cos(\vartheta)}{4} \Bigl ] 
\\
& C(\vartheta,\infty) = \frac{c_1 c_2 \varepsilon_1^t \varepsilon_2^t}{4}.
\end{split}
\end{equation}

\noindent
where $\varepsilon_i^t = \varepsilon_i^\uparrow + \varepsilon_i^\downarrow$.
Following the definition of Eq. (\ref{eq2}) and assuming identical polarizers ($\varepsilon_1^\downarrow
\varepsilon_2^\uparrow=\varepsilon_1^\uparrow\varepsilon_2^\downarrow$), we derive

\begin{equation} 
\begin{split}
P_{12}(\vartheta) & = {1 \over 4} \Big [1 - \frac{2\varepsilon_1^\downarrow \varepsilon_2^\uparrow}
{\varepsilon_1^t \varepsilon_2^t} \cos(\vartheta)\Big ], {\rm ~~for~~}
0^\circ \le \vartheta \le 90^\circ {\rm and}
\\ 
P_{12}(\vartheta) & = {1 \over 4} \Big [1 - \frac{\varepsilon_1^\uparrow \varepsilon_2^\uparrow + 
\varepsilon_1^\downarrow \varepsilon_2^\downarrow}
{\varepsilon_1^t \varepsilon_2^t} \cos(\vartheta) \Big ], {\rm ~~for~~}
90^\circ \le \vartheta \le 180^\circ 
\label{eq6}
\end{split}
\end{equation}

\noindent
 The results obtained by 
using Eq. (\ref{eq6}) are shown in Fig. \ref{fig3} (full line) as a function of $\vartheta$. Clearly 
the theoretical prediction of the quantum theory is in agreement with the experimental points, 
within the errors of the present experiment.

\begin{figure}%[hb]
  \centering
  \includegraphics[width=9cm]{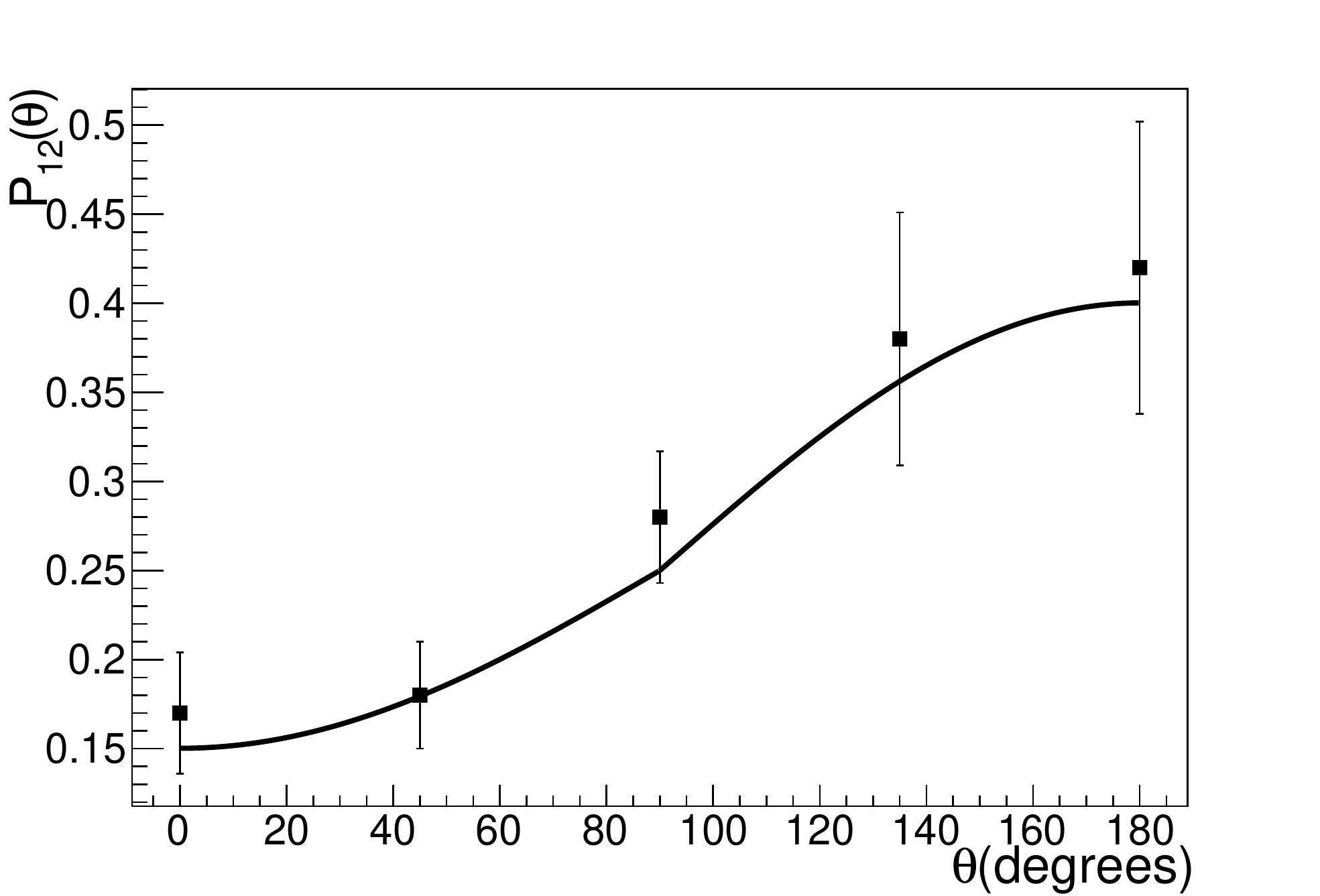}
  \caption{$P_{12}(\vartheta)$ as a function of the angle $\vartheta$ between the polarization 
  axes.  Two 4-pixel rows with relative separation 0.4 cm, and $\Delta$ = 150 ns. 
  Full line: quantum prediction (Eq. (\ref{eq6}))}
  \label{fig3}
\end{figure}

Recall also that for an analogous experiment which might be performed with ideal instrumentation 
and with a source emitting only neutron pairs in the singlet state, the quantum prediction
substituting Eq. (\ref{eq6}) is :

\begin{equation}
P^{\rm ideal}_{12} (\vartheta) = \langle\psi_s|Q^\downarrow_1Q^\downarrow_2|\psi_s\rangle = \frac{[(1-\cos(\vartheta)]}{4} , 
\label{eq7}
\end{equation}

\noindent and that this simple sinusoidal form of the ideal prediction (as other simple sinusoidal 
forms of quantum predictions, in other cases) is the sole origin for the violations of the CH 
inequality \cite{ballantine}.

Let us summarize the most significant experimental features of the present measurement:
(1) from the highly monochromatic and collimated beam of thermal neutrons produced at the IN10 
beam of the ILL, by utilizing the anti-bunching effect of the neutrons in the triplet states, the two 
members of a pair in the singlet state could be detected at small time separation as a function of 
the angle $\vartheta$ between the two polarizers;
(2) the effects on the data analysis due to the presence of undesirable and uncertain instrumental 
origin were greatly reduced by measuring $P_{12}(\vartheta)$ from the ratio of the coincidence rate at 
small time separation to the coincidence rate at large time separation ($t_2 - t_1 \gg \tau_c$).
   
The measurement seems adequate to confirm  the quantum-theoretical prediction (\ref{eq6}) 
which takes account of the real limitations of the present experiment, and, by extrapolation, the 
theoretical prediction (\ref{eq7}) for an ideal experiment performed with perfect instrumentation. 

We wish to remark explicitly that the efficiency of our polarizers was insufficient for a decisive 
confrontation of the experimental data with the Clauser-Horne (CH) inequality, confrontation 
which requires very high efficiency of all of the components of the apparatus. 
We also note that such limitation prevent our from performing loophole-free measurements. 
On the other hand, we consider our result as a helpful contribution to this issue. Actually, the 
confirmation of quantum prediction (\ref{eq6}) could but be reinforced by experiments performed 
with greater efficiency and statistical precision, as it was shown in the optical polarization experiments, 
and also commented by Bell \cite{bell2}: ".. it is hard for me to believe that quantum mechanics works 
so nicely for inefficient practical set-up and is yet going to fail badly when sufficient refinements 
are made ..."
Therefore, our results seem adequate to verify the predictions of the quantum theory for massive 
particles in the spin singlet state, and, although by extrapolation from the real measurement, 
to confirm violations of the CH inequality.

We are grateful to P. Facchi and S. Pascazio for enlightening discussions 
and we acknowledge the Institut Laue Langevin of Grenoble for the beam time, which enabled 
the realization of the experiment.

\end{document}